\documentclass{article}
\usepackage{graphicx}
\usepackage[small]{subfigure,epsfig}
\usepackage {amsmath} \usepackage{amssymb}
\usepackage{cite}
\usepackage[onehalfspacing]{setspace}
\usepackage{indentfirst}

\usepackage{geometry} 
\begin{document}
\title{Application of the Kudryashov method for finding exact solutions of the high order nonlinear evolution equations}
\author{Pavel N. Ryabov\footnote{E-mail: pnryabov@mephi.ru}, Dmitry I. Sinelshchikov, Mark B. Kochanov}
\date{Department of Applied Mathematics, National  Research Nuclear University MEPHI, 31 Kashirskoe
Shosse, 115409 Moscow, Russian Federation}

\maketitle
\begin{abstract}
The application of the Kudryashov method for finding exact solutions
of the high  order nonlinear evolution equations is considered. Some
classes of solitary wave solutions for the families of nonlinear
evolution equations of fifth, sixth and seventh order are obtained.
The efficiency of the Kudryashov method for finding exact solutions
of the high order nonlinear evolution equations is demonstrated.
\end{abstract}



\section{Introduction}

The powerful and effective method for finding exact  solutions of
nonlinear ordinary differential equations was proposed in work
\cite{Kudryashov_1988}. In works \cite{Kudryashov1990, Kudryashov91,
Kudryashov1993} author applied this method to construct the exact
solutions of the nonlinear nonintegrable equations. The first
modification of this method was presented in work
\cite{Kudryashov_book}. The final and the most successful
modification of this method was proposed in \cite{KudryashovArx}.
Thus, we refer this method as the Kudryashov method. The Kudryashov
method allow us in the straightforward manner to construct solitary
wave solutions for a wide class of nonlinear ordinary differential
equations. The main idea of the Kudryashov method is to use special
form of the singularity manifold in the truncation method
\cite{Kudryashov_1988,Kudryashov1990,Kudryashov91,Kudryashov1993,Kudryashov_book}.
This approach allows us to reduce the problem of constructing exact
solutions to solving the overdetermined system of algebraic
equations.  The main advantage of the Kudryashov method is that we
can more effectively construct exact solutions of high order
nonlinear evolution equations in comparison with other methods for
finding exact solutions
\cite{Parkes1994,Parkes01,Malfliet,Fan2000,Biswas,Vitanov, Loginova}.

Using this method in works
\cite{Kudryashov_1988,Kudryashov1990,Kudryashov91} exact  solutions
of the generalized Kuramoto--Sivashinsky equation, the
Burgers--Korteweg--de Vries eqution, the Bretherton eqaution and the
Kawahara equation were obtained. Traveling wave solutions for class
of third order nonlinear evolution equations were constructed in
\cite{Ryabov} with help of the Kudryashov method. The Kudryashov
method was used for constructing  traveling wave solutions of
several nonlinear evolution equation in works \cite{Kabir1,
Gepreel2011} as well.

The aim of this work is to demonstrate efficiency of the Kudryashov
method for  finding exact solitary wave solutions of high order
nonlinear evolution equations. For this purpose we consider three
families of nonlinear evolution equations of fifth, six and seven
order. This families of equations have some applications in physics
and other fields of science. We present classes of solitary wave
solutions for considered families of nonlinear evolution equations.

This work is organized as follows. In the next section we give brief
description of  the Kudryashov method algorithm. In the sections 3,4
and 5 we construct solitary wave solutions for the families of
nonlinear evolution equations of fifth, six and seven order
respectively. In the last section we summarize and discuss our
results.

\section{Method applied}
The aim of this section is to present the algorithm of the
Kudryashov method for finding exact solutions of the nonlinear
evolution equations. To reach this purpose we will follow the works
\cite{KudryashovArx,  Kudryashov_book, Ryabov}.

Let us consider the nonlinear partial differential equation in the form
\begin{equation}
E_1[u_{t},u_{x},\ldots,x,t]=0
\label{ch2:eq1.1}
\end{equation}

Using the following ansatz
\begin{equation}
u(x,t)=y(z), \ z=kx-\omega t.
\label{ch2:eq: tr_wave}
\end{equation}
from Eq. \eqref{ch2:eq1.1} we obtain the ordinary nonlinear differential equation
\begin{equation}
E_2[-\omega y_z,ky_{z},k^2 y_{zz}, k^3 y_{zzz}, \ldots]=0.
\label{ch2:eq1.3}
\end{equation}

Now we show how one could obtain the exact solution of  the Eq.
\eqref{ch2:eq1.3} using the approach by Kudryashov. This method
consist of the following steps \cite{KudryashovArx, Kudryashov_book,
Ryabov}.

\emph{The first step. Determination of the dominant terms.}

To find dominant terms we substitute
\begin{equation}
y=z^p,
\end{equation}
into all terms of Eq. \eqref{ch2:eq1.3}. Then we compare degrees of
all terms in Eq. \eqref{ch2:eq1.3}  and choose two or more with the
smallest degree. The minimum value of $p$ define the pole of Eq.
\eqref{ch2:eq1.3} solution and we denote it as $N$. We have to point
out that method can be applied when $N$ is integer. If the value $N$
is noninteger one can transform the equation studied and repeat the
procedure.

\emph{The second step. The solution structure}.

We look for exact solution of Eq. \eqref{ch2:eq1.3} in the form
\begin{equation}\label{eq2.5}
y=a_0+a_1Q(z)+a_2(z)Q(z)^2+...+a_{N}Q(z)^{N},
\end{equation}
where  $a_i$ - unknown constants, $Q(z)$ is the following function
\begin{equation}
Q(z)=\frac{1}{1+e^z}. \label{eq2.6}
\end{equation}

This function satisfies to the first order ordinary differential
equation
\begin{equation}\label{DiffEq}
Q_z=Q^2-Q.
\end{equation}

The Eq. \eqref{DiffEq} is necessary to calculate the derivatives of
function $y(z)$.

\emph{The third step. Derivatives calculation}.

We should calculate all derivatives of function $y$. One can do it using the computer
algebra systems Maple or Mathematica. As an example we consider following case: \\

\emph{The derivatives of function $y(z)$ in the case of $N=2$ can be written in the form}
\begin{equation}
\begin{gathered}
y=a_0+a_1Q+a_2Q^2,\\
y_{z}=-a_1Q+(a_1-2a_2)Q^2+2a_2Q^3,\\
y_{zz}=a_1Q+(4a_2-3a_1)Q^2+(2a_1-10a_2)Q^3+6a_2Q^4.
\end{gathered}
\label{eq2.7}
\end{equation}

The relations \eqref{eq2.7} can be generalized for any value of $N$.
Differentiating the expression \eqref{eq2.5} with respect to $z$ and
taking into account \eqref{DiffEq} we have
\begin{equation}\label{GenYdiff}
\begin{gathered}
y_z=\sum_{i=1}^{N} a_i i (Q-1)Q^i, \\
y_{zz}=\sum_{i=1}^{N} a_i i ((i+1)Q^2-(2i+1)Q+i)Q^i.
\end{gathered}
\end{equation}

The high order derivatives of function $y(z)$ can be found in works
\cite{KudryashovArx, Kudryashov_book}.


\emph{The fourth step. Defining the values of unknown parameters}.

We substitute expressions \eqref{GenYdiff} in Eq. \eqref{ch2:eq1.1}.
After it we have take into account \eqref{eq2.5}. Thus Eq.
\eqref{ch2:eq1.1} takes the form
\begin{equation}
P[Q(z)]=0,
\end{equation}
where $P[Q(z)]$ - is a polynomial of function $Q(z)$. Then we
collect all items with the same powers of function $Q(z)$ and equate
this expressions equal to zero. As a result we obtain algebraic
system of equations. Solving this system we get the values of
unknown parameters.

The Kudryashov method is a very powerful method for finding exact
solutions of the nonlinear differential equations. It has a set of
advantages. They are:
\begin{enumerate}
\item The first step is not necessary, because all redundant terms in \eqref{eq2.5} becomes equal to zero when we start to define unknown parameters from the system of algebraic equations;
\item We construct the solution as a set of $Q$-blocks. The function $Q$ does not contain any parameters. It is very comfortable, because all of them is in equation;
\item Given method can be easily programmed in Maple or Mathematica, because we use the substitutions \eqref{eq2.5} and \eqref{GenYdiff} in Eq. \eqref{ch2:eq1.3} it takes the polynomial
form;
\item This method is powerful and effective even if we construct the exact solutions of the high order nonlinear evolution equations.
\item It easy to show that tanh, coth, (G'/G) – methods and Kudryashov method can be reduced to each other \cite{KudryashovArx}. Moreover the Kudryashov method gives the same results as an Exp-function method. However it is well known that Exp-function method can not be applied for the equations of hight order;
\end{enumerate}

Let us give several examples to demonstrate it's efficiency.

\section{Exact solitary wave solutions of the fifth order evolution equation}
As an example let us consider the fifth order nonlinear evolution
equation in the form
\begin{equation}\label{ch3:eq1}
\begin{gathered}
u_t+uu_x+10uu_{xxx}+20u_xu_{xx}+30u^2u_{x}+\alpha u_{xx}+\beta
u_{xxx}+\\+\gamma u_{xxxx}+u_{xxxxx}=0.
\end{gathered}
\end{equation}

Eq. \eqref{ch3:eq1} is new and does not present in the periodic literature. However it is an
interesting equation because this equation consist of the fifth order Korteweg-de Vries equation
with additional dispersive and dissipative terms. The equation \eqref{ch3:eq1} can arise in the
physical applications when we consider the wave processes in active dispersive-dissipative media.

Using the traveling waves \eqref{ch2:eq: tr_wave} we have
\begin{equation}\label{ch3:eq2}
\begin{gathered}
-wy_z+kyy_z+10k^3yy_{zzz}+20k^3y_zy_{zz}+30ky^2y_{z}+\alpha k^2
y_{zz}+\beta k^3 y_{zzz}+\\+\gamma k^4 y_{zzzz}+ k^5 y_{zzzzz}=0.
\end{gathered}
\end{equation}

The pole of the Eq. \eqref{ch3:eq2} is equal to $N=2$, thus we look
for exact solution in the form
\begin{equation}\label{ch3:eq3}
y=a_0+a_1Q+a_2Q^2,
\end{equation}
where $a_0, a_1$ and $a_2$ -- are unknown constants.

Substituting \eqref{GenYdiff} in Eq. \eqref{ch3:eq2} and taking into
account \eqref{ch3:eq3} we obtain the polynomial of function $Q(z)$.
Collecting all terms with the same power of function $Q(z)$ and
equate this expressions to zero we obtain the system of algebraic
equations. Solving this system we find that solution of Eq.
\eqref{ch3:eq2} exists only in eight cases. They are
\begin{equation}\label{ch3:eq4}
\begin{gathered}
k_{(1,2)}=\pm\frac{\gamma}{21}, \quad
k_{(3,4)}=\pm\frac{3\gamma}{7},  \quad
k_{(5,6)}=\pm\frac{1}{84}\sqrt{988\gamma^2+2646\beta-441}, \\
k_{(7,8)}=\pm\frac{1}{35\gamma}\sqrt{5\gamma(972\gamma^3+343\alpha)}.
\end{gathered}
\end{equation}

However in the case of $k=k_{(5,6)}$ and $k=k_{(7,8)}$ the
parameters and solution presentation is very cumbersome and we do
not give them in the present manuscript.

In the case of $k=k_{(1,2)}$ and $k=k_{(3,4)}$ the values of
parameters $\alpha, \beta, w, a_0, a_1, a_2$ are
\begin{equation}\label{ch3:eq5}
\begin{gathered}
k=\frac{\gamma}{21}, \quad
\alpha=\frac{5}{84}\gamma - \frac{5}{14}\gamma\beta - \frac {44}{441}\gamma^{3}, \quad \\
a_0=\frac{\beta}{20}+\frac{16\gamma^2}{2205}-\frac{1}{40}, \quad
a_1=\frac{4}{147}\gamma^{2}, \quad
a_2=-\frac{2}{147}\gamma^{2}, \\
\omega=\frac {\gamma\left( 9472\gamma^{4}-10416\gamma^{2}-9261-
37044\beta+62496\beta\gamma^{2}+111132\beta^{2} \right)}{31116960} ,
\end{gathered}
\end{equation}

\begin{equation}\label{ch3:eq6}
\begin{gathered}
k=-\frac{\gamma}{21}, \quad
\alpha=\frac{5}{84}\gamma - \frac{5}{14}\gamma\beta - \frac {44}{441}\gamma^{3}, \quad \\
a_0=\frac{\beta}{20}+\frac{46\gamma^2}{2205}-\frac{1}{40}, \quad
a_1=0, \quad
a_2=-\frac{2}{147}\gamma^{2}, \\
\omega=-\frac {\gamma\left( 9472\gamma^{4}-10416\gamma^{2}-9261-
37044\beta+62496\beta\gamma^{2}+111132\beta^{2} \right)}{31116960} ,
\end{gathered}
\end{equation}

\begin{equation}\label{ch3:eq7}
\begin{gathered}
k=\frac{3\gamma}{7}, \quad
\beta=\frac{1}{6}-\frac{18}{7}\gamma^2, \\
a_0=\frac{7\alpha}{150\gamma} - \frac{1}{60} +
\frac{222}{1225}\gamma^{2}, \quad a_1=0, \quad
a_2=-\frac{18}{49}\gamma^{2}, \\
\omega=\frac{67228\alpha^2-7056\alpha\gamma^3+347328\gamma^6-8575\gamma^2}{2401000\gamma},
\end{gathered}
\end{equation}

\begin{equation}\label{ch3:eq8}
\begin{gathered}
k=-\frac{3\gamma}{7}, \quad
\beta=\frac{1}{6}-\frac{18}{7}\gamma^2, \\
a_0=\frac{7\alpha}{150\gamma} - \frac{1}{60} -
\frac{228}{1225}\gamma^{2}, \quad a_1=\frac {36}{49}\gamma^{2},
\quad
a_2=-\frac {18}{49}\gamma^{2}, \\
\omega=-\frac{67228\alpha^2-7056\gamma^3\alpha+347328\gamma^6-8575\gamma^2}{2401000\gamma},
\end{gathered}
\end{equation}

The solutions of the Eq. \eqref{ch3:eq2} which corresponds to the
relations \eqref{ch3:eq5}--\eqref{ch3:eq8} are
\begin{gather}
y(z)=a_0+\frac{2}{147}\gamma^2\left[2-Q(z)\right]Q(z),\label{ch3:eq9}\\
y(z)=a_0-\frac{2\gamma^2}{147}Q(z)^2,\label{ch3:eq10}\\
y(z)=a_0-\frac{18\gamma^2}{49}Q(z)^2,\label{ch3:eq11}\\
y(z)=a_0+\frac{18\gamma^2}{49}\left[2-Q(z)\right]Q(z).\label{ch3:eq12}
\end{gather}

\begin{figure}[!htb]
\center
\includegraphics[width=70 mm]{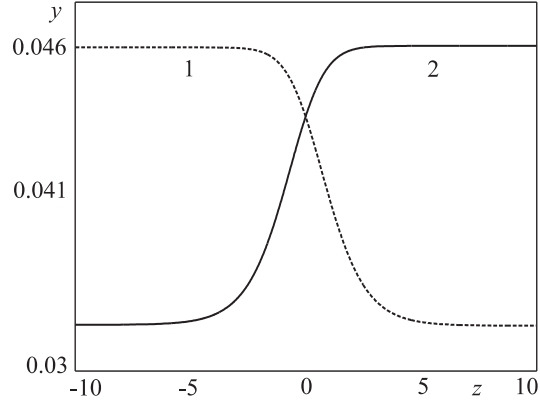}
\caption{{\fontshape{sl}\selectfont Exact solutions \eqref{ch3:eq9}
and  \eqref{ch3:eq10} -- 1, 2 respectively  at
$\beta=\gamma=1$.}}\label{fig1}
\end{figure}

Graphical presentation of solutions \eqref{ch3:eq9},
\eqref{ch3:eq10} is shown on Fig. \ref{fig1}.

The solutions \eqref{ch3:eq11}, \eqref{ch3:eq12} have the same
structure so we have decided not to picture them.

\section{Exact solitary waves of the sixth order evolution equation}
Let us consider the following equation
\begin{equation}\label{ch4:eq1}
u_t+uu_x+\alpha u_{xx}+\beta u_{xxxx}+\gamma u_{xxxxxx}=0.
\end{equation}

This equation was  proposed in work \cite{Nikolaevskii} for
describing the longitudinal seismic waves in a viscoelastic medium.
Moreover this equation is used for modeling the ''soft'' -- type of
the turbulence \cite{Tribelsky}. Also this equation describes the
chemical reactions in reaction-diffusion systems \cite{Tanaka}. The
numerical modeling of the wave processes describing by Eq.
\eqref{ch4:eq1} was performed in work \cite{Kudryashov2007}. Exact
solutions of this equation was obtained in work
\cite{Kudryashov2004}.

Taking into account the traveling waves \eqref{ch2:eq: tr_wave} in
Eq. \eqref{ch4:eq1} we obtain
\begin{equation}\label{ch4:eq2}
-\omega y_z+kyy_z+\alpha k^2 y_{zz}+\beta k^4 y_{zzzz}+\gamma k^6 y_{zzzzzz}=0.
\end{equation}
Integrating the Eq. \eqref{ch4:eq2} we have
\begin{equation}\label{ch4:eq3}
C_1-\omega y+k\frac{y^2}{2}+\alpha k^2 y_{z}+\beta k^4 y_{zzz}+\gamma k^6 y_{zzzzz}=0.
\end{equation}

Dominant terms of the Eq. \eqref{ch4:eq3} are $k^6y_{zzzzz}, ky^2/2$. Thus, the pole order of the Eq. \eqref{ch4:eq3} solution is $N = 5$. So we look for solution in the form
\begin{equation}\label{ch4:eq4}
y(z)=a_0+a_1Q+a_2Q^2+a_3Q^3+a_4Q^4+a_5Q^5,
\end{equation}
where $a_0$ - $a_5$ -- constants to be determined.

Using the \eqref{GenYdiff} in Eq. \eqref{ch4:eq3} and taking into
account ansatz \eqref{ch4:eq4} we obtain a system of algebraic
equations. Solving this system we find four real families of unknown
parameters. However we give only three because one of them is
cumbersome. The following families are

\begin{equation}\label{ch4:eq5}
\begin{gathered}
\alpha=-\frac{114}{5}k^2\beta, \quad \gamma=-\frac{\beta}{55k^2}, \quad C_1=\frac{\omega^2}{2k}-\frac{127008k^7\beta^2}{121}, \\
a_0=\frac{\omega}{k}-\frac{504k^3\beta}{11}, \quad a_1=a_2=0, \quad a_3=\frac{10080k^3\beta}{11},\\
a_4=-\frac{15120k^3\beta}{11}, \quad a_5=\frac{6048k^3\beta}{11}
\end{gathered}
\end{equation}

\begin{equation}\label{ch4:eq6}
\begin{gathered}
\alpha=\frac{219}{44}k^2\beta, \quad \gamma=-\frac{5}{44}\frac{\beta}{k^2}, \quad C_1=\frac{\omega^2}{2k}-\frac{99225k^7\beta^2}{968},\\
a_0=\frac{\omega}{k}+\frac{315k^3\beta}{22}, \quad a_1=0, \quad a_2=-\frac{19845k^3\beta}{11}, \quad a_3=6930k^3\beta,\\
a_4=-\frac{94500k^3\beta}{11}, \quad a_5=\frac{37800k^3\beta}{11}
\end{gathered}
\end{equation}


\begin{equation}\label{ch4:eq7}
\begin{gathered}
\alpha=-\frac{3259}{110}\beta k^2,\quad \gamma=-\frac{1}{110}\frac{\beta}{k^2}, \quad C_1=\frac{\omega^2}{2k}-\frac{321489k^7\beta^2}{242}, \\
a_0=\frac{\omega}{k}-\frac{567k^3\beta}{11}, \quad a_1=0, \quad a_2=\frac{1890k^3\beta}{11}, \quad a_3=\frac{3780k^3\beta}{11},\\
a_4=-\frac{7560k^3\beta}{11}, \quad a_5=\frac{3024k^3\beta}{11}
\end{gathered}
\end{equation}

From relations \eqref{ch4:eq5}, \eqref{ch4:eq6} and \eqref{ch4:eq7}
we see that the real solution of Eq. \eqref{ch4:eq3} exists in the
case of
\begin{equation}\label{ch4:eq8}
\begin{gathered}
k^2=-\frac{5}{114}\frac{\alpha}{\beta}, \quad \beta^2=\frac{275}{114}\alpha\gamma, \\
k^2=\frac{44}{219}\frac{\alpha}{\beta}, \quad \beta^2=-\frac{1095}{1936}\alpha\gamma, \\
k^2=-\frac{110}{3259}\frac{\alpha}{\beta}, \quad \beta^2=\frac{3259}{12100}\alpha\gamma.
\end{gathered}
\end{equation}

The solutions of Eq. \eqref{ch4:eq3} which corresponds to the relations \eqref{ch4:eq5}, \eqref{ch4:eq6} and \eqref{ch4:eq7} takes the form
\begin{equation}\label{ch4:eq9}
y(z)=\frac{\omega}{k}+\frac{\beta k^3}{11}\left(-504+10080Q(z)^3-15120Q(z)^4+6048Q(z)^5\right),
\end{equation}

\begin{equation}\label{ch4:eq10}
\begin{gathered}
y(z)=\frac{\omega}{k}+\frac{\beta k^3}{11}\left(\frac{315}{2}-19845Q(z)^2+76230Q(z)^3-94500Q(z)^4+\right.\\ \left.+37800Q(z)^5\right),
\end{gathered}
\end{equation}

\begin{equation}\label{ch4:eq11}
\begin{gathered}
y(z)=\frac{\omega}{k}+\frac{\beta k^3}{11}\left(-567+1890Q(z)^2+3780Q(z)^3-7560Q(z)^4+\right.\\ \left.+3024Q(z)^5\right).
\end{gathered}
\end{equation}

\begin{figure}[!htb]
\center
\includegraphics[width=70.0 mm]{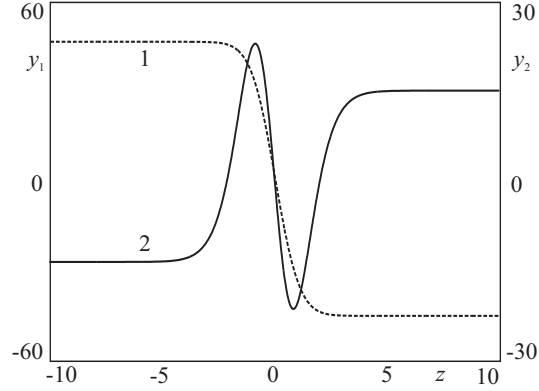}
\caption{{\fontshape{sl}\selectfont Exact solutions \eqref{ch4:eq9}
and  \eqref{ch4:eq10} -- 1, 2 respectively  at
$\beta=\omega=k=1$.}}\label{fig2}
\end{figure}

The solutions \eqref{ch4:eq9}, \eqref{ch4:eq10} of the Eq.
\eqref{ch4:eq1} is presented on Fig. \ref{fig2}. The third solution
has the form of the kink as well as \eqref{ch4:eq9}.

\section{Exact solitary waves of the seventh order evolution equation}
Let us consider the nonlinear evolution equation of seventh order
\begin{equation}\label{ch5:eq1}
u_t+u^nu_x+\alpha u_{xxx} +\beta u_{xxxxx}+\gamma u_{xxxxxxx}=0,
\end{equation}
where $n=1,2,3$.

For the first time this equation was considered in work
\cite{KudryashovArx} in the case $n=2$. We generalizes the results
of this work.

Taking the traveling wave ansatz $u(x, t) = y(z), z = kx - wt$  into account from Eq.\eqref{ch5:eq1} we have
\begin{equation}\label{ch5:eq2}
-\omega y_z+ky^ny_z+\alpha k^3 y_{zzz}+\beta k^5 y_{zzzzz}+\gamma k^7 y_{zzzzzzz}=0.
\end{equation}

Integrating the Eq. \eqref{ch5:eq2} with respect to variable $z$ we obtain
\begin{equation}\label{ch5:eq3}
C_1-\omega y+k\frac{y^{n+1}}{n+1}+\alpha k^3 y_{zz}+\beta k^5 y_{zzzz}+\gamma k^7 y_{zzzzzz}=0.
\end{equation}

In the case of $n=1$ we have to look exact solution of Eq. \eqref{ch5:eq3} in the form
\begin{equation}\label{ch5:eq4}
y(z)=a_0+a_1Q+a_2Q^2+a_3Q^3+a_4Q^4+a_5Q^5+a_6Q^6,
\end{equation}
because the pole order of Eq. \eqref{ch5:eq3} solution is $N = 6$.

Substituting \eqref{GenYdiff} in Eq. \eqref{ch5:eq3} with $n = 1$
and taking \eqref{ch5:eq4} into account we found the following
solutions
\begin{equation}\label{ch5:eq5}
\begin{gathered}
y(z)=\frac{\omega}{k}+\gamma k^6 \left(-7100+166320Q(z)^2+332640Q(z)^3-\right. \\ \left. -1829520Q(z)^4+1995840Q(z)^5-665280Q(z)^6\right),
\end{gathered}
\end{equation}

\begin{equation}\label{ch5:eq6}
\begin{gathered}
y(z)=\frac{\omega}{k}+\gamma k^6 \left(-3600+665280Q(z)^3-1995840Q(z)^4\right. \\ \left. +1995840Q(z)^5-665280Q(z)^6\right),
\end{gathered}
\end{equation}

The corresponding parameters of Eq. \eqref{ch5:eq3} and \eqref{ch5:eq4} are
\begin{equation}\label{ch5:eq7}
\begin{gathered}
\beta=-100k^2\gamma, \quad \alpha=2159k^4\gamma, \quad
C_1=\frac{\omega^2}{2k}-25205000k^{13}\gamma^2,\\
a_0=\frac{\omega}{k}-7100k^6\gamma, \quad a_1=0, \quad a_2=166320k^6\gamma, \quad a_3=332640k^6\gamma, \\ a_4=-1829520k^6\gamma, \quad
a_5=1995840k^6\gamma, \quad a_6=-665280k^6\gamma,
\end{gathered}
\end{equation}

\begin{equation}\label{ch5:eq8}
\begin{gathered}
\beta=-50k^2\gamma, \quad \alpha=769k^4\gamma, \quad
C_1=\frac{\omega^2}{2k}-6480000k^{13}\gamma^2, \\
a_0=\frac{\omega}{k}-3600k^6\gamma, \quad a_1=a_2=0, \quad a_3=665280k^6\gamma, \\
a_4=-1995840k^6\gamma, \quad a_5=-a_4, \quad a_6=-a_3,
\end{gathered}
\end{equation}
where $\beta/\gamma<0$. Thus, the solutions \eqref{ch5:eq5} and \eqref{ch5:eq6} exist when
\begin{equation}
\begin{gathered}
k^2=-\frac{\beta}{100\gamma}, \quad \beta^2=\frac{1000}{2159}\alpha\gamma,\\
k^2=-\frac{\beta}{50\gamma}, \quad \beta^2=\frac{2500}{769}\alpha\gamma.
\end{gathered}
\end{equation}

\begin{figure}[!htb]
\center
\includegraphics[width=70.0 mm]{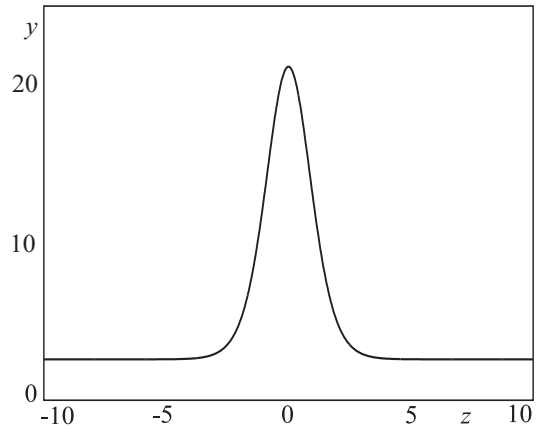}
\caption{{\fontshape{sl}\selectfont Exact solution \eqref{ch5:eq5}
at $\omega=10, \gamma=0.001, k=1$.}}\label{fig3}
\end{figure}

The illustration of \eqref{ch5:eq5} is given on Fig. \ref{fig3}.

The solution \eqref{ch5:eq6} has the same form.

Let us consider the case of $n = 2$. The pole order of Eq.
\eqref{ch5:eq3} solution in that case is equal to $N = 3$. Thus, we
look for exact solution of Eq. \eqref{ch5:eq3} in the form
\begin{equation}\label{ch5:eq9}
y(z)=a_0+a_1Q+a_2Q^2+a_3Q^3,
\end{equation}
where $a_0, a_1, a_2, a_3$ – the constants to be determined.

Taking into account \eqref{GenYdiff} and relations \eqref{ch5:eq9}
in Eq. \eqref{ch5:eq3} we have the solution in the form

\begin{equation}\label{ch5:eq10}
y(z)=\pm6\sqrt{-105\gamma}k^3\left(1-6Q(z)^2+4Q(z)^3\right),
\end{equation}
where $\gamma<0$.

Values of $\beta, \alpha, w, C_1$ are defined by relations
\begin{equation}\label{ch5:eq11}
\begin{gathered}
\beta=-83k^2\gamma, \quad \alpha=946k^4\gamma, \quad
\quad w=-1260\gamma k^7, \quad C_1=0, \\
a_0=\pm6\sqrt{-105\gamma}k^3, \quad a_1=0, \quad a_2=\mp36\sqrt{-105\gamma}k^3, \\ a_3=\pm24\sqrt{-105\gamma}k^3,
\end{gathered}
\end{equation}

So, the solution \eqref{ch5:eq10} exist when the wave number satisfies to equivalence
\begin{equation}
k^2=-\frac{\beta}{83\gamma}, \quad \beta^2=\frac{6889}{946}\alpha\gamma.
\end{equation}

\begin{figure}[!htb]
\center
\includegraphics[width=70.0 mm]{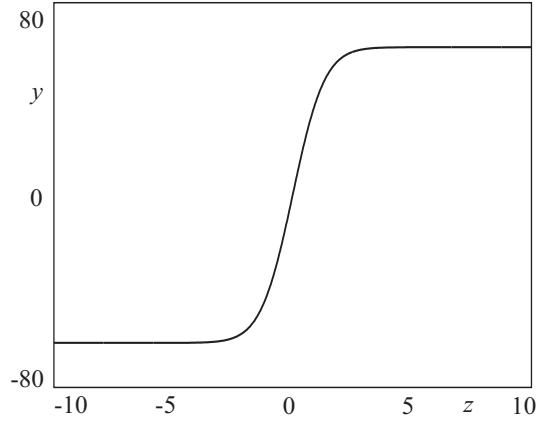}
\caption{{\fontshape{sl}\selectfont Exact solution \eqref{ch5:eq10}
at $\gamma=-1, k=1$.}}\label{fig4}
\end{figure}

The solution \eqref{ch5:eq10} is a kink, see Fig. \ref{fig4}.

In the case of $n = 3$ the pole order of the Eq. \eqref{ch5:eq3}
solution is $N = 2$. So, we use the following ansatz
\begin{equation}\label{ch5:eq12}
y(z)=a_0+a_1Q+a_2Q^2,
\end{equation}
where $a_0, a_1, a_2$ -- unknown constants.

Then we substitute \eqref{GenYdiff} and \eqref{ch5:eq12} in Eq.
(5.3) with $n = 3$. Collecting all terms with the same power of
function $Q(z)$ and equate them to zero we obtain the system of
algebraic equations on unknown parameters. Solving this system we
have
\begin{equation}\label{ch5:eq13}
\begin{gathered}
\alpha=14k^4\gamma+\frac{5\beta^2}{28\gamma}, \quad
w=\frac{k}{1176\gamma^2}\left(3920k^6\gamma^3 -1764 \beta \gamma^2 k^4 - 5\beta^3\right), \\
C_1=\frac{\sqrt[3]{315}}{65856}\frac{k}{\gamma^{8/3}}\left(5\beta^4-137200k^8\gamma^4+9408k^6\gamma^3\beta+392k^4\gamma^2\beta^2\right), \\
a_0=-\frac{\sqrt[3]{315}}{42}\frac{\beta}{\gamma^{2/3}}-\frac{\sqrt[3]{315\gamma}k^2}{3}, \quad a_1=4\sqrt[3]{315\gamma}k^2, \quad a_2=-a_1.
\end{gathered}
\end{equation}

\begin{figure}[!htb]
\center
\includegraphics[width=70.0 mm]{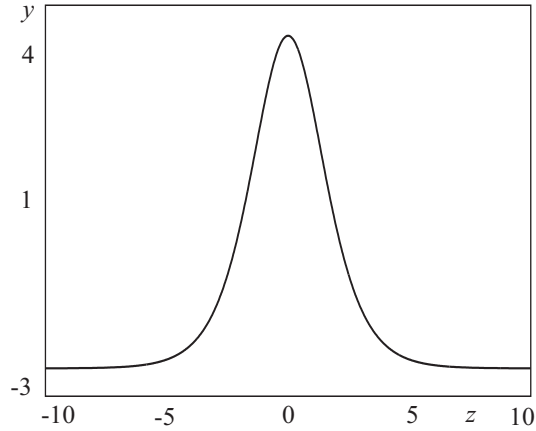}
\caption{{\fontshape{sl}\selectfont Exact solution \eqref{ch5:eq14}
at $\beta=\gamma=k=1$.}}\label{fig5}
\end{figure}

The solution of Eq. \eqref{ch5:eq3} with $n = 3$ which corresponds to \eqref{ch5:eq13} takes the form
\begin{equation}\label{ch5:eq14}
y(z)=-\frac{\sqrt[3]{315}}{42}\frac{\beta}{\gamma^{2/3}}-\frac{\sqrt[3]{315\gamma}k^2}{3}\left(1-12Q(z)+12Q(z)^2\right).
\end{equation}
This \eqref{ch5:eq14} has the form of the traveling wave (Fig.
\ref{fig5}).

\section{Conclusion}
In this work we have demonstrated efficiency of the Kudryashov
method for finding  exact solutions of high order nonlinear
evolution equations. We have obtained solitary wave solutions for
the three families of nonlinear evolution equations of fifth, six
and seven orders. Graphical representation of these exact solution
is presented. We believe that some of exact solutions obtained in
this work are new.

\section{Acknowledgements}
This work was supported by the federal target programm ''Research
and scientific-pedagogical personnel of innovation in Russia'' \ on
2009-2011.

\end{document}